\documentclass[conference]{IEEEtran}
\IEEEoverridecommandlockouts
\usepackage{cite}
\usepackage{amsmath,amssymb,amsfonts}
\usepackage{algorithmic}
\usepackage[dvipdfmx]{graphicx}
\usepackage{textcomp}
\usepackage{xcolor}
\usepackage{multirow}
\usepackage{comment}

\def\BibTeX{{\rm B\kern-.05em{\sc i\kern-.025em b}\kern-.08em
    T\kern-.1667em\lower.7ex\hbox{E}\kern-.125emX}}
\begin{document}
\title{RVCoreP-32IC: A high-performance RISC-V soft processor with an efficient fetch unit supporting the compressed instructions}
\author{\IEEEauthorblockN{Takuto Kanamori\IEEEauthorrefmark{1},
Hiromu Miyazaki\IEEEauthorrefmark{1} and
Kenji Kise\IEEEauthorrefmark{1}}
\IEEEauthorblockA{\IEEEauthorrefmark{1}School of Computing\\
Tokyo Institute of Technology, Tokyo, Japan\\
Email: \{kanamori, miyazaki, kise\}@arch.cs.titech.ac.jp}}

\maketitle

\begin{abstract}
In this paper, we propose a high-performance RISC-V soft processor with an efficient fetch unit supporting the compressed instructions targeting on FPGA.
The compressed instruction extension in RISC-V can reduce the program size by about 25\%.
But it needs a complicated logic for the instruction fetch unit and has a significant impact on performance.

We propose an instruction fetch unit that supports
the compressed instructions while exhibiting high performance.
Furthermore, we propose a RISC-V soft processor using this unit.
We implement this proposed processor in Verilog HDL and verify
the behavior using Verilog simulation and an actual Xilinx Atrix-7 FPGA board.
We compare the results of some benchmarks and the amount of hardware with related works.
DMIPS, CoreMark value, and Embench value of the proposed processor achieved 42.5\%, 41.1\%
and 21.3\% higher performance than the related work, respectively.
\end{abstract}


\section{Introduction}

A computer system with a soft processor like MicroBlaze\cite{MicroBlaze} and Nios II\cite{Nios} is implemented on an FPGA, which is used in various fields. RISC-V is receiving attention as an ISA (Instruction Set Architecture) adopted by soft processors.

RISC-V is a RISC (Reduced Instruction Set Computer) based ISA designed to be universal and extensible based on lessons learned from past instruction sets. A processor designer can select the required instruction sets according to the application requirements. As an extended instruction set that can be added to RV32I, that is the basic integer instruction set, "M" for multiplication and division instructions, "F" for single-precision floating-point instructions, "A" for atomic instructions needed to support the modern OS instructions are defined.

The "C" extension is a compressed instructions extension and replaces some frequently occurring 32-bit instructions with 16-bit instructions. So it is possible to prevent the code size from growing, which is a weak point of a RISC-based ISA. The other RISC-based ISAs that supports two types of instruction length have existed. The feature of the compressed instructions compared with these is no mode switching to support the compressed instructions, and all instructions are aligned on 16-bit boundaries instead of 32-bit boundaries. When adopting an existing processor to the compressed instructions, the performance will drop significantly if the instruction fetch unit is not changed appropriately.

We propose an efficient instruction fetch unit that supports the compressed instructions and a soft processor called RVCoreP-32IC or RVP-c, in short, using that unit. We implement this proposed processor in Verilog HDL and compare the results of some benchmarks and the amount of hardware with related works.

\section{Related works}
\begin{figure*}[htb]
  \begin{center}
    \includegraphics[clip,width=\linewidth]{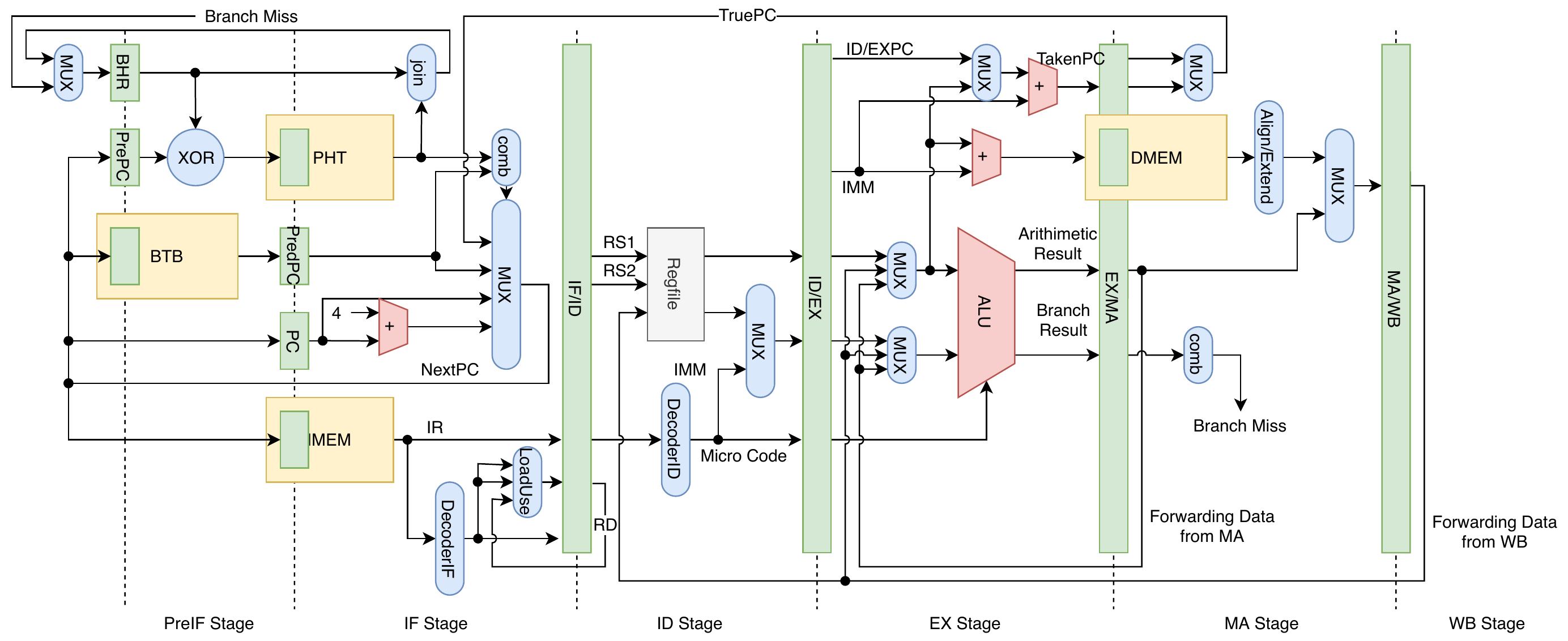}
  \end{center}
  \caption{The block diagram of the baseline processor RVCoreP-32I}
  \label{fig:arch-baseline-proc}
\end{figure*}

RVCoreP\cite{miyazaki2020rvcorep} is a RISC-V soft processor with a 5-stage pipeline targeting FPGA created by Miyazaki et al. It supports RV32I and is written by Verilog HDL.

We define RVCoreP as a baseline and modify it for supporting the compressed instructions. The version that does not support compressed instructions is named RVCoreP-32I(RVP).

Figure \ref{fig:arch-baseline-proc} shows the block diagram of RVCoreP-32I. The green squares are the register updated in synchronization with the rising edge of the clock signal, the yellow squares are the module consisted of Block RAM, the gray square is a module composed of LUT RAM that performs reading asynchronously and writing synchronously with the rising edge of the clock, the red modules are adders and ALUs, and other blue modules are combinational circuits.

RVP uses gshare\cite{gshare} for branch prediction. Pattern History Table (PHT) and Branch Target Buffer (BTB) are implemented using Block RAM. The branch prediction mechanism is pipelined to improve the operating frequency. \cite{matsuibpred}

\begin{figure}[tb]
    \begin{center}
      \includegraphics[clip,width=\columnwidth]{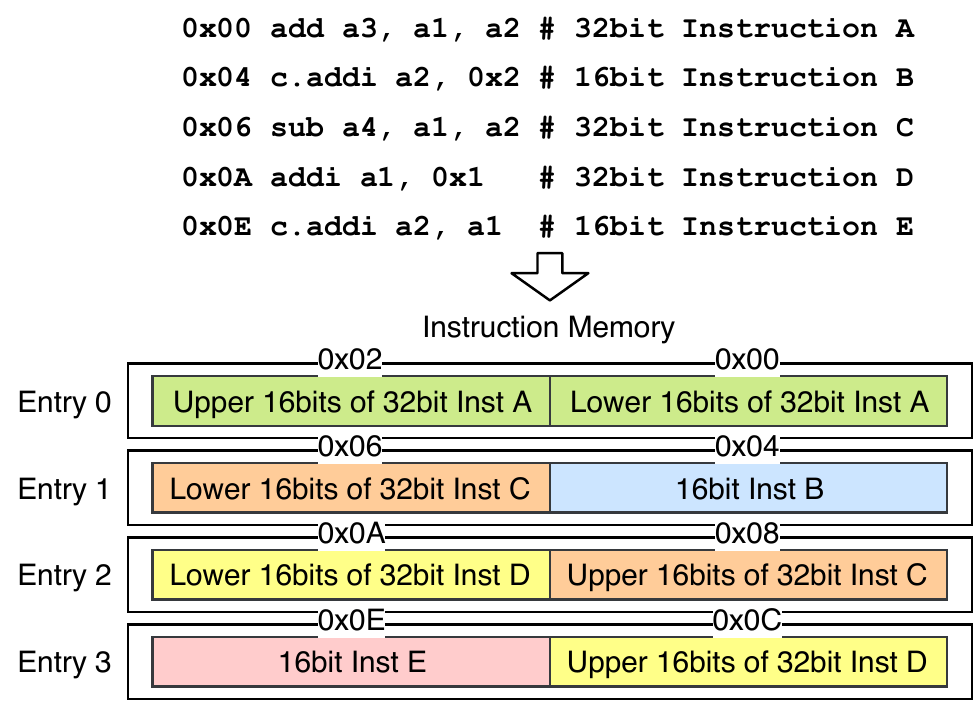}
    \end{center}
    \caption{The figure of 32-bit wide instruction memory with mixed 32-bit and 16-bit instructions}
    \label{fig:design-imem32}
\end{figure}

The most important point to consider when supporting the compressed instructions is that 32-bit instructions and 16-bit instructions coexist in the instruction memory and are arranged without gaps. Figure \ref{fig:design-imem32} shows the situation where 32-bit instructions and 16-bit instructions are placed in the 32-bit wide instruction memory. Hereafter, the width of memory that can be accessed using one I/O port is defined as an entry. In this figure, one entry consists of 32-bit. For explanation, addresses are assigned every 16-bit in the memory.

For example, the 32-bit \textit{Inst C} located at addresses 0x06 and 0x08 shown in orange is divided into two entries and placed both Entry 1 and Entry 2. Therefore, in order to fetch \textit{Inst C}, it is necessary to access both Entry 1 and Entry 2. The IPC (Instruction Per Cycle) of the processor is significantly decreased if the instruction memory is accessed in two steps in order to fetch a 32-bit instruction crossing a boundary of entries like a \textit{Inst C}.

\begin{figure}[tb]
    \begin{center}
      \includegraphics[clip,width=\columnwidth]{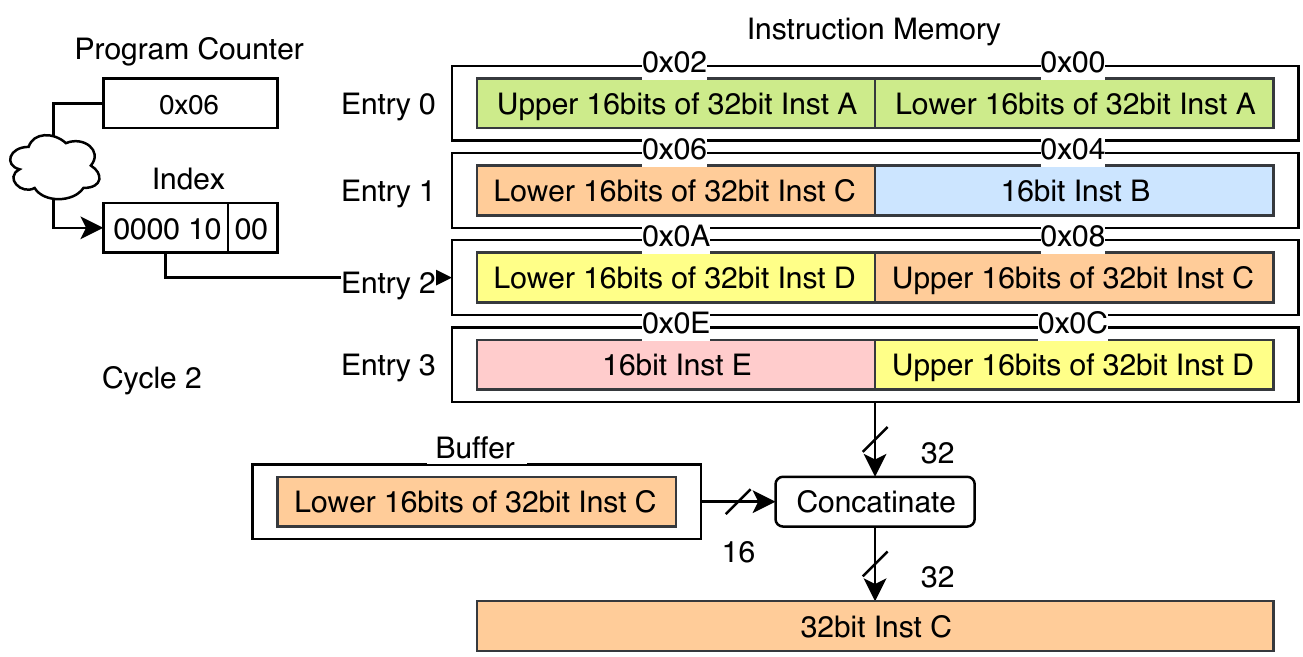}
    \end{center}
    \caption{The figure of the buffering instruction fetch architecture}
    \label{fig:design-imem32-buf}
\end{figure}

The method of buffering the data fetched from the instruction memory is often used to avoid this problem in the instruction fetch unit supporting the compressed instructions. Figure \ref{fig:design-imem32-buf} shows the situation where \textit{Inst C} is fetched in the buffering instruction fetch unit. Each processor in PULP Platform\cite{pulp} and Syntacore's SCR1\cite{scr1} use FIFO buffers, and VexRiscv\cite{vexriscv} uses a minimum 16-bit buffer.

However, it is necessary to delete the value of the buffer when the branch instruction is taken. Therefore, when branching to a 32-bit instruction that is not aligned on a 32-bit boundary, the lower 16-bit of the instruction can be fetched, but the upper 16-bit of the instruction cannot be fetched with the same access. Hereafter referred to as \textit{fetch miss}. This results in a lower IPC.

\begin{figure}[tb]
    \begin{center}
      \includegraphics[clip,width=\columnwidth]{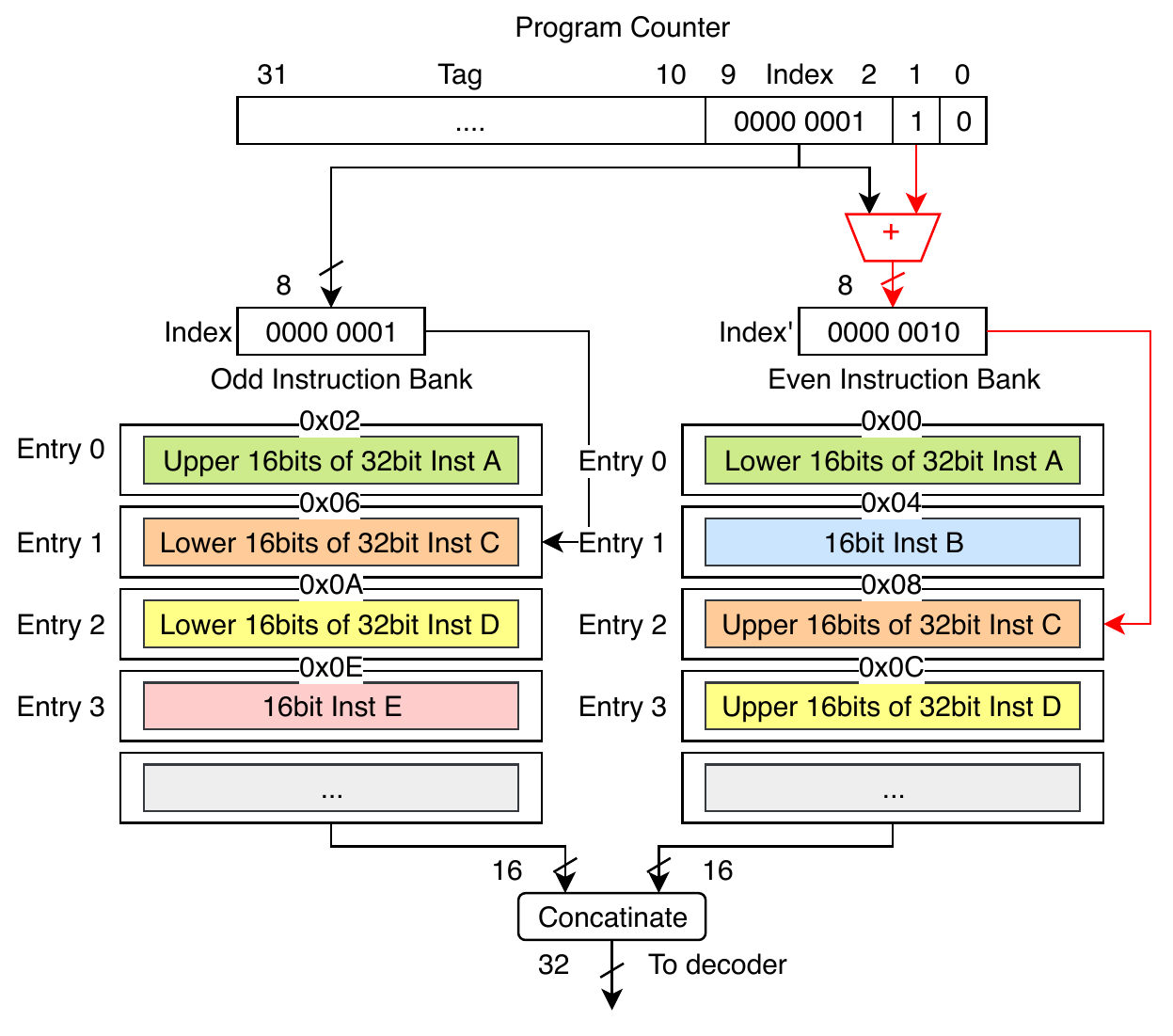}
    \end{center}
    \caption{The figure of Gray's instruction cache structure minimized and optimized for RISC-V}
    \label{fig:related-I-cache}
\end{figure}

Gray's instruction cache structure \cite{vondran2002efficient} can efficiently fetch variable-length instructions of CISC (Complex Instruction Set Computer) and VLIW (Very Long Instruction Word) based ISA. Figure \ref{fig:related-I-cache} shows the situation where \textit{Inst C} is fetched in Gray's instruction cache structure minimized and optimized for RISC-V. The entries in this figure consist of 16-bit.

In this instruction cache structure, each instruction is placed into Odd Instruction Bank and Even Instruction Bank according to the address. Since each bank is accessed in parallel, any instruction can be efficiently fetched without accessing in two steps.

After the address of one bank is calculated, the address of the other bank is calculated by adding to the address. So the red path in the figure is the critical path of the instruction fetch structure. In this structure, a circuit that adds only one is used to alleviate the critical path. However, the circuit is targeted for implementation as an ASIC and is not expected to be implemented on FPGA.

\section{Proposed method}
\subsection{The instruction fetch unit}

The proposed instruction fetch unit has two program counters, and it accesses two entries simultaneously, similar to Gray's instruction cache structure. But unlike Gray's instruction cache structure, it does not divide the instruction memory into two. Since the Block RAM for most FPGAs has two I/O ports, the proposed unit accesses two entries of one instruction memory composed of block RAM at the same time.

The width of the instruction memory of the baseline processor is 32 bits. If the two entries in this memory are always accessed, at least 32-bit of data will be wasted. So we change the width of the instruction memory to 16-bit. After that, one entry in the memory is 16-bit.

\begin{figure}[tb]
    \begin{center}
      \includegraphics[clip,width=\columnwidth]{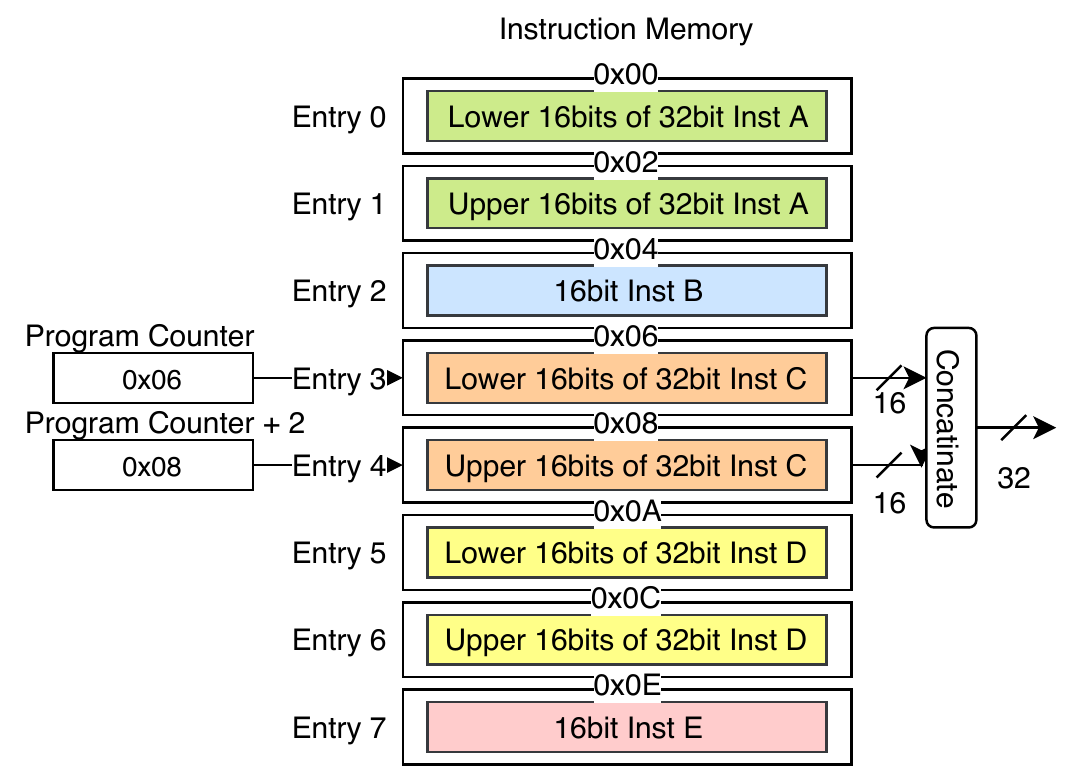}
    \end{center}
    \caption{The figure of instruction fetch with two program counters into 16-bit wide instruction memory}
    \label{fig:design-imem16}
\end{figure}

Figure \ref{fig:design-imem16} shows the situation that \textit{Inst C} located in the 16-bit width instruction memory is fetched using two program counters. \textit{Program Counter} (PC) in the figure is the original program counter, \textit{Program Counter + 2} (PC\_2) in the figure is the program counter that points to the next entry pointed to by the original program counter. The proposed unit consists of these two program counters to access two consecutive entries always. Therefore, even if a branch destination is a 32-bit instruction that is not aligned on a 32-bit boundary, the proposed unit can fetch the instruction in one cycle.

\begin{figure}[bt]
    \begin{center}
      \includegraphics[clip,width=\columnwidth]{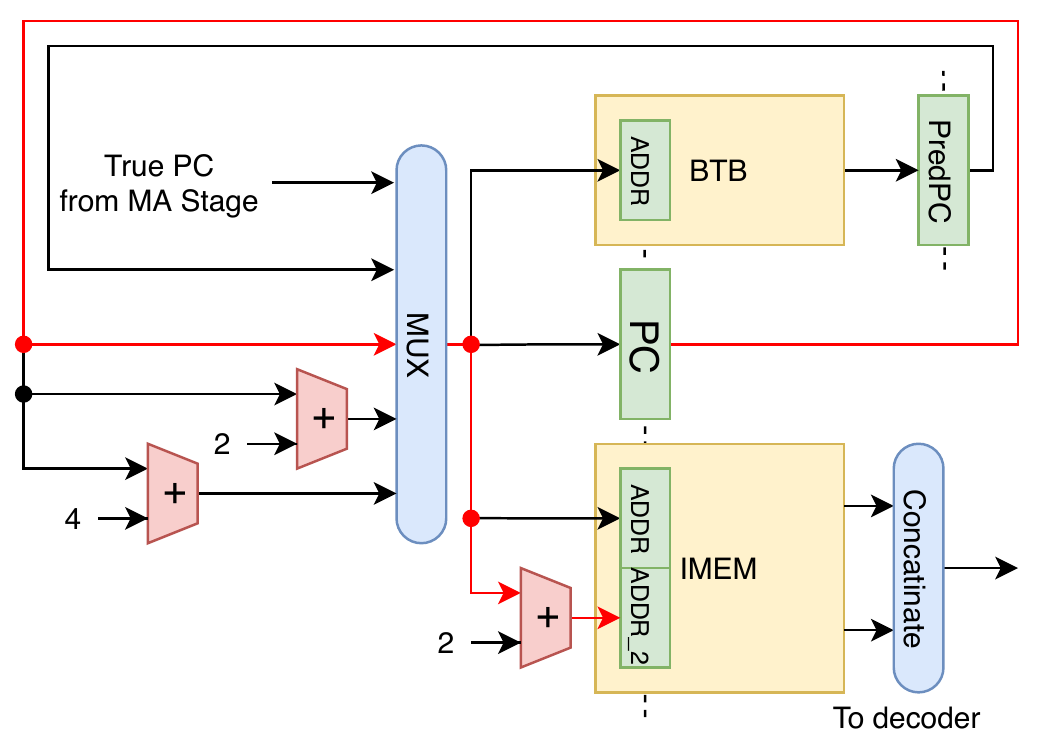}
    \end{center}
    \caption{The block diagram of a simple implementation of instruction fetch with access to two entries}
    \label{fig:arch-slow-fetch}
\end{figure}

Figure \ref{fig:arch-slow-fetch} shows a block diagram of the simple instruction fetch unit using the instruction memory shown in Figure \ref{fig:design-imem16}. It is simply implemented to access two entries in the instruction fetch unit of the baseline processor. After calculating the value of PC like Gray's instruction cache, this simple implementation adds 2 to that value to calculate the value of PC\_2.

However, when this unit is applied to the baseline processor, the red path in the figure becomes a critical path. So the operating frequency is significantly decreased. Since the critical path of the baseline processor is the path that calculates the value of PC, the addition of circuits to this path should be minimized.

\begin{figure}[tb]
    \begin{center}
      \includegraphics[clip,width=\columnwidth]{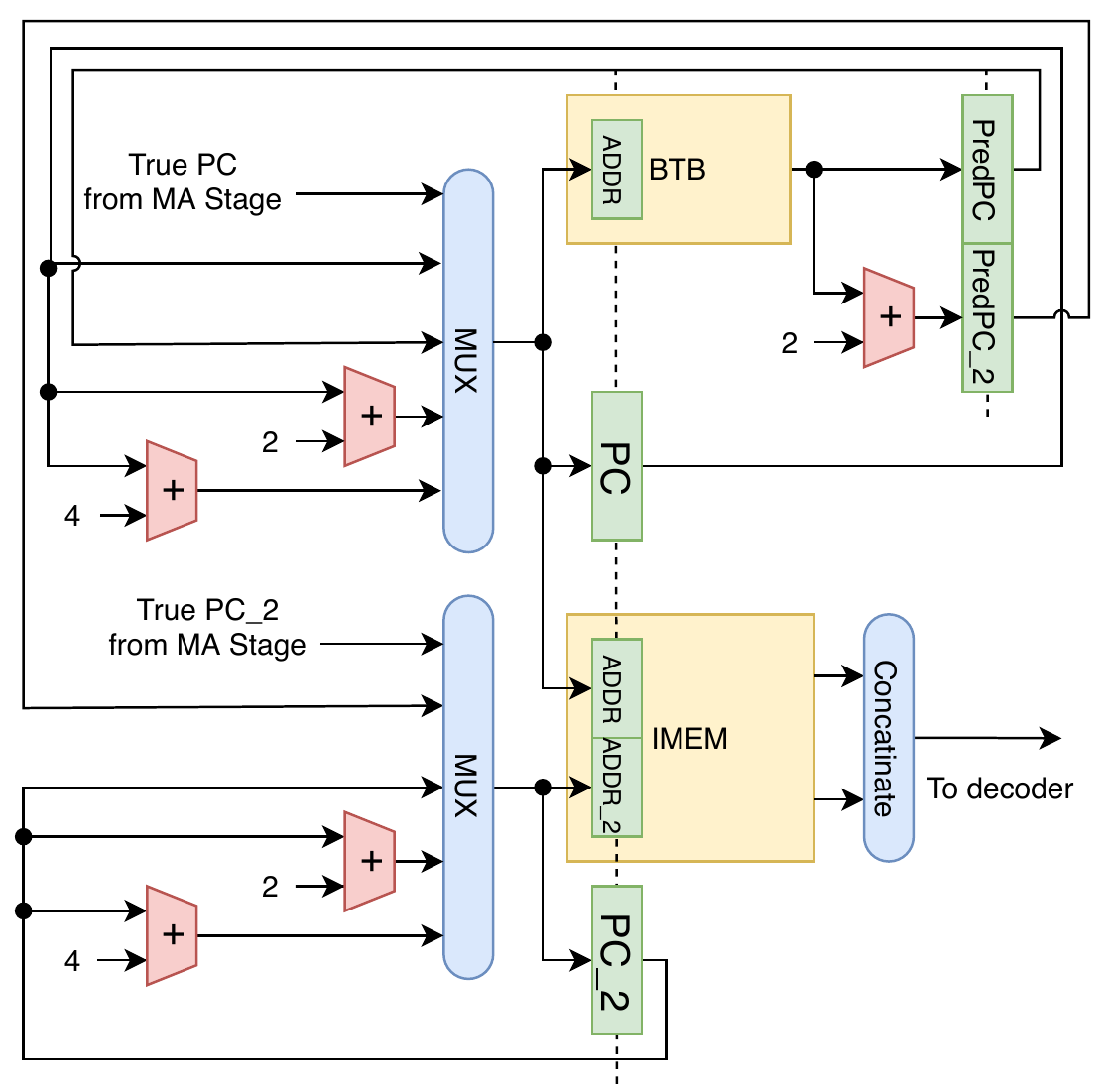}
    \end{center}
    \caption{The block diagram of the proposed instruction fetch unit}
    \label{fig:arch-fetch}
\end{figure}

Figure \ref{fig:arch-fetch} shows the block diagram of the proposed instruction fetch unit. The value of the next cycle PC is selected from 5 candidates in the IF stage. The proposed unit pre-calculates the values obtained by adding 2 to all these five candidates and selects the value of PC\_2 at the same timing as the selection of the value of PC.

The value of next cycle PC is any of these five candidates in the RVCoreP-32IC, a value obtained by adding 2 or 4 to the value of the current PC (PC+2,PC+4), the value of the current PC for to stall, the address of the branch prediction destination read from BTB (PredPC) or the address of the branch target to correct the branch misprediction sent from the MA stage (TruePC).

The value obtained by adding 2 to the value of PC, PC+2 and PC+4 can be calculated by replicating the selection logic for PC and adders, adding register saving the value of PC\_2.

The value obtained by adding 2 to the value of PredPC (PredPC\_2) is calculated by placing an adder immediately after BTB. In the baseline processor, the value of PredPC is written to the register immediately after being read from BTB, so adding an adder to this path does not become a critical path.

\begin{figure}[tb]
    \begin{center}
      \includegraphics[clip,width=\columnwidth]{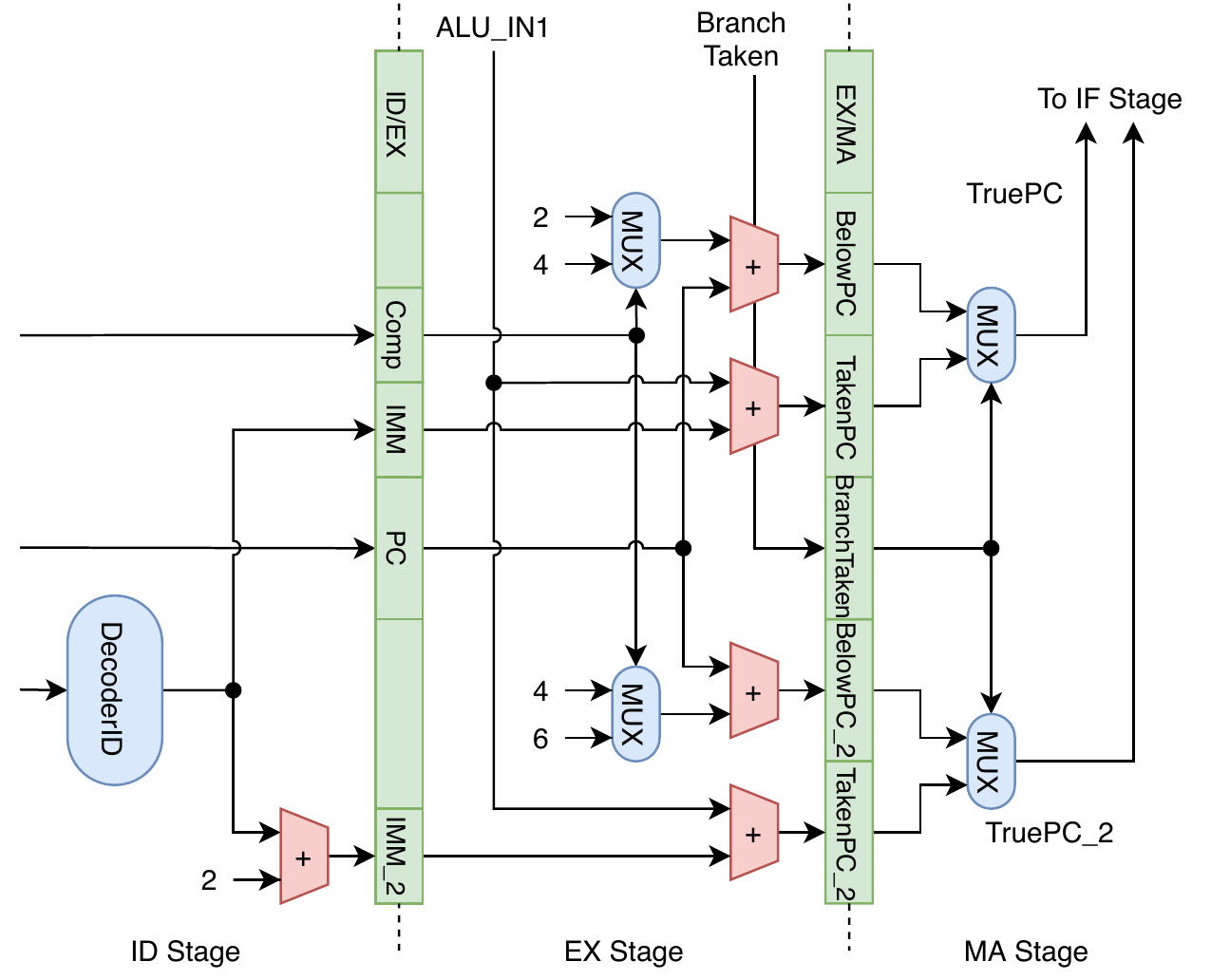}
    \end{center}
    \caption{The block diagram of the circuit calculating TruePC\_2 in the pipeline}
    \label{fig:arch-TruePC}
\end{figure}

Figure \ref{fig:arch-TruePC} shows a pipelined circuit that calculates the value that is obtained by adding 2 to the value of TruePC (TruePC\_2). There are two candidates for TruePC. When a branch instruction is predicted as taken and the branch prediction is missed, it is the address of the next instruction on the instruction memory(BelowPC). On the other hand, when a branch instruction is predicted as not taken and the branch prediction is missed, it is the address of the correct branch destination(TakenPC). We define the values obtained by adding 2 to these two kinds of values as BelowPC\_2 and TakenPC\_2. These values are calculated in the processor pipeline.

When the branch instruction is a 16-bit instruction, BelowPC is obtained by adding 2 to the address and BelowPC\_2 is obtained by addign 4 to the address. Samely, when the branch instruction is a 32-bit, BelowPC is obtained by adding 4 to the address and BelowPC\_2 is obtained by adding 6 to the address. In the EX stage, either 4 or 6 is selected by using the \textit{Comp} of the ID/EX pipeline register that indicates whether the branch instruction is a compressed instruction. And this value added to the address of the branch instruction.

TakenPC is the branch destination address. RISC-V branch instructions are divided into two types of Unconditional Jumps and Conditional Branches. The branch destination address of Unconditional Jumps is the value obtained by adding the decoded immediate value (IMM) to either the address of the branch instruction or the value of the operand register. In Conditional Branches, it is the value obtained by adding IMM to the address value of the branch instruction.

Therefore, the branch destination of the RISC-V branch instruction is either the value obtained by adding a branch instruction address to IMM or adding an operand register value to IMM. In order to calculate the values obtained by adding 2 to both candidates, the value obtained by adding 2 to IMM (IMM\_2) is calculated in advance. Since there is relatively a room in the path of generation IMM in the ID stage of the baseline processor, we place an adder after DecoderID to calculate IMM\_2. TakenPC\_2 is calculated parallelly with the normal branch destination in the EX stage. In the MA stage, BelowPC\_2 or TakenPC\_2 is selected and sent to the IF stage as TruePC\_2 by using the \textit{BranchTaken} of the EX/MA pipeline register that indicates whether branch is taken or not.

\subsection{The proposed RISC-V soft processor RVCoreP-32IC}

\begin{figure*}[htb]
  \begin{center}
    \includegraphics[clip,width=\linewidth]{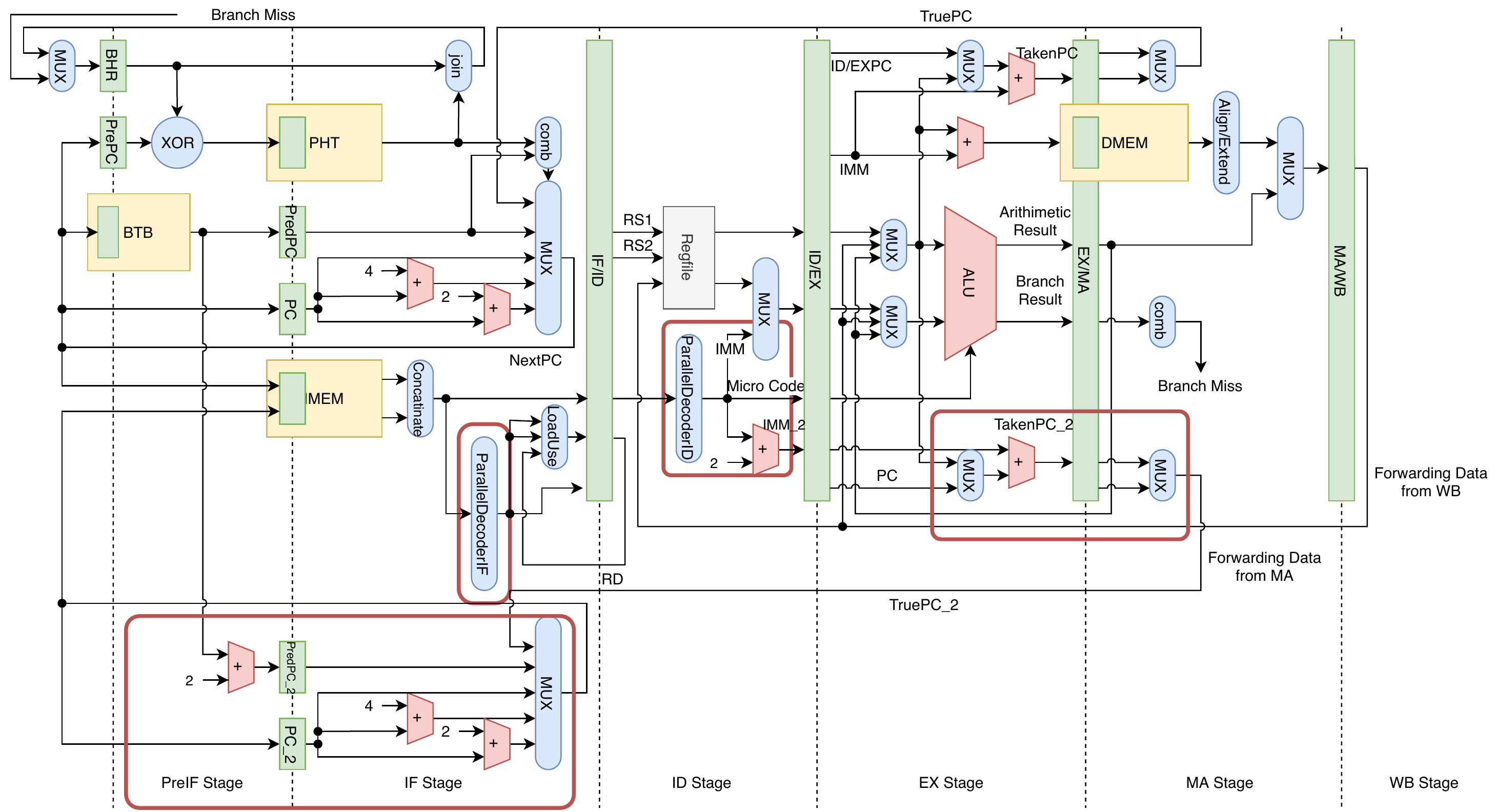}
  \end{center}
  \caption{The block diagram of the proposed processor RVCoreP-32IC supporting the compressed instructions}
  \label{fig:arch-proposal-proc}
\end{figure*}


Figure \ref{fig:arch-proposal-proc} shows the block diagram of the proposed processor RVCoreP-32IC. The parts surrounded by the red frame are the changes from the baseline processor RVCoreP-32I. In addition to the proposed instruction fetch unit, changes are made to the decoders and the control of the branch prediction mechanism.

To detect the dependency with the instruction at the IF stage and the load instruction at the ID stage, the decoders of RVP is divided into DecoderIF in the IF stage and DecoderID in the ID stage. If a circuit that decompresses the 16-bit instruction to the 32-bit instruction (\textit{Decompressor}) is placed before DecoderIF to support the compressed instructions in RVP, this path becomes the critical path. Therefore we implement ParallelDecoderIF of RVP-c in the IF stage, which decodes 16-bit instructions in parallel to 32-bit instructions without decompressing 16-bit instructions to avoid this problem.

In the ID stage, IMM\_2 is calculated after generating IMM with DecoderID. If a \textit{Decompressor} is added before the Decoder, this path becomes a critical path too. Therefore, we implement ParallelDecoderID similar to ParallelDecoderIF.

In the no cycle penalty pipelined branch prediction mechanism implemented in RVP, in order to generate the index used for writing to PHT and BTB, the address of the previous instruction on the instruction memory is used.

When it supports the compressed instructions, the address of the previous instruction is either the address obtained by minus 2 or 4 to the address of the branch instruction. RVP-c uses the information of the previous instruction in the pipeline register to determine the address.

However, if the previous instruction of the branch instruction in the pipeline is also a branch instruction and the previous branch instruction is predicted as taken, the previous instruction does not match the previous instruction in the instruction memory. In this case, writing to PHT and BTB should be prohibited.

\section{Verification and Evaluation}
\subsection{Verification}

We verified the implemented RTL code by Verilog simulation. A RISC-V processor simulator modeling a conservative multi-cycle processor named SimRV that we implemented in C++ is used as the reference model.

SimRV outputs the PC value, the executed instruction, and the 32 values stored in the register file when a RISC-V program binary is given. By executing the same binary using SimRV and Verilog simulation for our designed processors, log files of the same format can be output. We executed all benchmark binaries used in the evaluation described later and compared each log file. We have confirmed that their values in two log files match, and the programs are executing correctly.

In addition to the verification through simulations, we verified the behavior of the designed processor using an FPGA board. The same RISC-V program binary used for Verilog simulation is executed on the actual Xilinx Atrix-7 FPGA board, and we have confirmed that the ASCII character output of the execution results via a serial communication had matched to the correct result, and confirmed that the numbers of execution cycles and executed instructions are also matched.

\subsection{Evaluation environment}

We compare RVCoreP and VexRiscv that won the 2018 RISC-V SoftCPU Contest sponsored by the RISC-V Foundation \cite{cpucontest}. The source code of VexRiscv used for evaluation is published on GitHub, and the used version is \textit{Spinal-HDL/VexRiscv@ca228a3} committed on September 26, 2019. In the preliminary evaluation, the versions of VexRiscv with and without the branch prediction mechanism had the equivalent performance values obtained by multiplying the operating frequency and IPC, so we prepared each version.

The version supporting RV32IC of VexRiscv with the branch prediction mechanism is VR-bp-c, and the version without the branch prediction mechanism is VR-nobp-c. We set the \textit{compressedGen} parameter as \textit{true} to support the compressed instructions and the other parameters of VexRiscv to be as close as possible to RVCoreP.

VR-bp-c implements a bimodal branch predictor and a BTB. The prediction scheme of the proposal is a gshare branch predictor, which achieves higher prediction accuracy than the bimodal predictor of VR-bp-c.

In order to compare fairly between VR-bp-c and RVP-c, We set the size of the block RAM used in the branch prediction mechanism the same. In RVP-c, the number of BTB entries is 512, and the number of PHT entries for gshare is 8,192. They are implemented as 4KB block RAM in total. In VR-bp-c, we set the branch mechanism option \textit{DYNAMIC\_TARGET} in BranchPlugin and \textit{historyRamSizeLog2} parameter as 512 to implement as 4KB block RAM in total.

We ran multiple benchmarks on each processor to evaluate performance. The benchmark programs was compiled by RISC-V cross-compiler for RV32IC published in riscv-gnu-toolchain\cite{riscv-gnu-toolchain}. The version of the used compiler is 8.1.0, and the optimization used option is -O2.

We evaluate the operating frequency and the hardware resources utilization targeting Digilent Nexys 4 DDR\cite{nexys4ddr} equipped with xc7a100tcsg324-1 of Xilinx Artix-7 FPGA family. We used Xilinx Vivado 2017.2 as the design tool.

We used the \textit{Flow\_PerfOptimaized\_high} strategy for logic synthesis and \textit{Performance\_ExplorePostRoutePhysOpt} for placement and routing. We performed the logic synthesis and placement and routing by incrementally changing the clock cycle constraint in 5MHz. The highest frequency that satisfies the constraints is used as the operating frequency of the processor. For hardware resource evaluation, we used the result of placement and routing at the maximum operating frequency.

In order to demonstrate the effectiveness of the proposed fetch unit, we implemented RVP-c-buf adopted the buffering mechanism in fetch architecture. It is different from RVP-c in only fetch unit, and all of the other is the same.

\subsection{Evaluation Results}
\subsubsection{Operating frequency and hardware utilization in minimum configuration}

\begin{table}[htb]
  \begin{center}
    \caption{The evaluation results of operating frequency and hardware utilization where 4KB memories are used.}
    \begin{tabular}[t]{|l|r||r|r|r|r|} \hline
                & RVP & RVP-c & RVP-c-buf & VR-nobp-c & VR-bp-c \\ \hline
      Freq [MHz]& 185 & \bf165 & 135 & 165 & 130 \\ \hline
      LUTs      & 1,086 & 1,402 & 1,389 & 1,001 & 1,064 \\ \hline
      Registers & 777 & 931 & 844 & 563 & 673 \\ \hline
      Slices    & 411	& 524	& 456	& 310	& 363 \\ \hline
    \end{tabular}
    \label{tab:eval4kb}
  \end{center}
\end{table}

Table \ref{tab:eval4kb} shows the maximum operating frequency and hardware resource utilization of each processor where 4KB instruction memory and data memory are used. The placement and routing were performed using only one clock region of the FPGA to stabilize the operating frequency of the evaluated system.

RVP-c consumes 29.0\%, 19.8\%, and 27.5\% more LUTs, registers, and slices than RVP, respectively. The operating frequency of RVP-c has dropped from 185MHz to 165MHz because the number of candidates for selecting the PC value has increased, which is the critical path of RVP. RVP-c consumes 31.8\%, 38.3\%, and 44.6\% more LUTs, registers, and slices than VR-bp-c, respectively. However, it achieves the same operating frequency as VR-nobp-c, which has no branch prediction mechanism.

VexRiscv implements \textit{Decompressor} at the IF stage because it has only decoding logic that supports 32-bit instructions. We analyzed the critical path of VexRiscv. As a result, in VR-nobp-c, we confirmed the path that includes \textit{Decompressor} became a critical path and caused a significant decrease in operating frequency. In VR-bp-c, the path for calculating the PC value, including the branch prediction mechanism, became a critical path. On the other hand, RVP-c performs the decoding of 16-bit instructions in parallel to the 32-bit instructions, thus minimizing the effect of decoding 16-bit instructions on the operating frequency. As RVP, the critical path is the path in the branch prediction mechanism.

By comparing RVP-c and RVP-c-buf, it can be seen that the proposed fetch unit consumes more resources and achieves a higher operating frequency than the buffering mechanism. Since the critical path of RVP-c-buf contains the control logic for the buffering mechanism, it is necessary to add one pipeline stage to improve the operating frequency.

\subsubsection{The results of Dhrystone and CoreMark}

\begin{figure*}[htb]
  \begin{center}
    \includegraphics[clip,width=\linewidth]{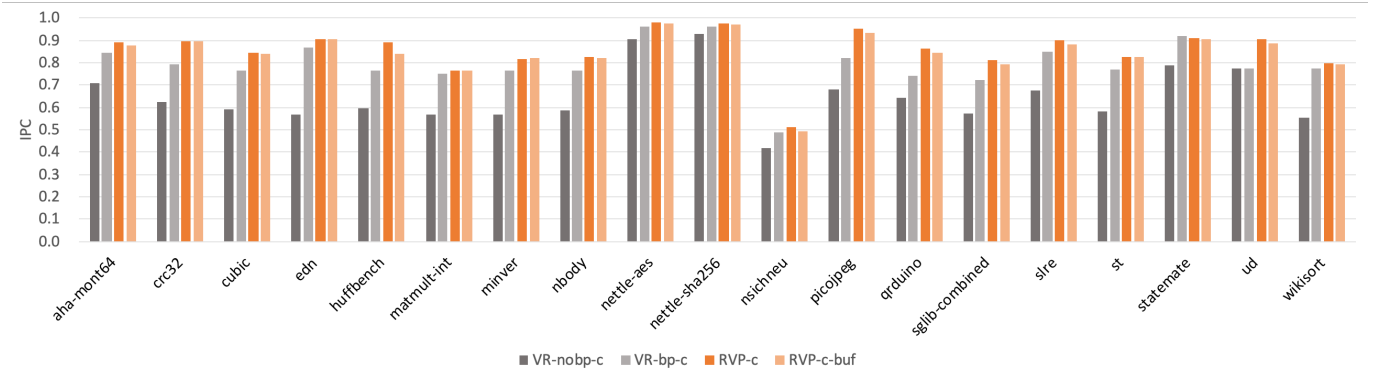}
  \end{center}
  \caption{The graphs of the IPC of each processor executing Embench where 64KB memories are used.}
  \label{fig:graph-embenchc}
\end{figure*}

\begin{table}[hbt]
  \begin{center}
    \caption{The evaluation results of operating frequency, Dhrystone MIPS and CoreMark value where 32KB memories are used.}
    \begin{tabular}[t]{|l|r|r|r|r|} \hline
                     & RVP-c  & RVP-c-buf & VR-nobp-c & VR-bp-c \\ \hline
      Freq[MHz]      & 150    & 125    & 145    & 120    \\ \hline
      DMIPS          & \bf172.4  & 139.1  & 116.4  & 121.0  \\ \hline
      DMIPS/MHz      & 1.149  & 1.113  & 0.803  & 1.009  \\ \hline
      DMIPS ratio    & 1.425  & 1.150  & 0.962  & 1.000  \\ \hline
      CoreMark       & \bf168.3  & 138.2  & 114.5  & 119.2  \\ \hline
      CoreMark/MHz   & 1.122  & 1.105  & 0.789  & 0.994  \\ \hline
      CoreMark ratio & 1.412  & 1.159  & 0.961  & 1.000  \\ \hline
    \end{tabular}
    \label{tab:eval32kb}
  \end{center}
\end{table}

Table \ref{tab:eval32kb} shows the operating frequency and the result of executing Dhrystone\cite{Dhrystone} and CoreMark\cite{coremark} of each processor where 32KB instruction memory and data memory are used. It is the minimum size that can store the benchmark programs, so we set the memory size 32KB. Note that each processor does not have a timer function, so the values in the table are calculated from the number of all calculation cycles and the number of executed instructions. The original value in the console output is higher than these values.

The source code of Dhrystone is published as riscv-tests\cite{riscvtests}. The option \textit{NUMBER\_OF\_RUNS} for the number of loops was set to 10000. In this case, the number of executed instructions is 4,526,099. The source code of CoreMark is released for RISC-V\cite{riscvcoremark}. The option \textit{ITERATIONS} for the number of loops was set to 10. In this case, the number of executed instructions is 7,346,906.

The row of DMIPS ratio and CoreMark ratio is the value where VR-bp-c is normalized as 1. From this result, it can be seen that RVP-c achieves 42.5\% and 41.2\%  higher performance than VR-bp-c when executing Dhrystone and CoreMark, respectively. The values of DMIPS/MHz and CoreMark/MHz of RVP-c are higher than that of RVP-c-buf, which shows that the proposed fetch unit is more efficient than the buffering method. The difference in DMIPS/MHz and CoreMark/MHz values of RVP-c-buf and VR-bp-c is caused by the difference in branch prediction accuracy between gshare and bimodal.

\subsubsection{The results of Embench}
\begin{table}[hbt]
  \begin{center}
    \caption{The evaluation results of operating frequency, Embench average IPC where 64KB memories are used.}
    \begin{tabular}[t]{|l|r|r|r|r|} \hline
                & RVP-c & RVP-c-buf & VR-nobp-c & VR-bp-c \\ \hline
    Freq[MHz]   & 135 & 125 & 145 & 120 \\ \hline
    Average IPC & \bf0.857  & 0.846  & 0.649  & 0.795  \\ \hline
    Average hit rate & 0.788  & 0.798  & N/A & 0.779  \\ \hline
    Performance & 115.7  & 105.7  & 94.1 & 95.4  \\ \hline
    Performance ratio & \bf1.213  & 1.108  & 0.987  & 1.000 \\ \hline
    \end{tabular}
    \label{tab:eval64kb}
  \end{center}
\end{table}

Figure \ref{fig:graph-embenchc} shows the IPC of each processor executing each Embench program.\cite{embench} The dark-gray bars, the light-gray bars, the dark-orange bars and the light-orange bars are the IPC of VR-nobp-c ,VR-bp-c, RVP-c and RVP-c-buf, respectively. RVP-c achieves the highest IPC in most benchmark programs, but VR-bp-c in \textit{statemete} and RVP-c-buf in \textit{minver} achieve the highest IPC.

Table \ref{tab:eval64kb} shows the operating frequency where 64KB instruction memory and data memory are used, the evaluation results of average IPC and average branch accuracy obtained by Verilog simulation. It is the minimum size that can store the benchmark programs, so we set the memory size 64KB. The row of Performance in this table is the value obtained by multiplying the operating frequency by IPC, and the bottom row is the Performance value where VR-bp-c is normalized as 1. From this result, it can be seen that RVP-c achieves 21.3\% higher performance than VR-bp-c when executing Embench.

The branch prediction mechanism of RVP-c and RVP-c-buf has the same structure, but the branch prediction accuracy is different because the update timing of BTB and PHT may differ due to \textit{fetch miss}. Although RVP-buf-c achieves higher branch prediction accuracy, RVP-c achieves higher IPC because of \textit{fetch miss} in the buffering mechanism. This result shows that the fetch unit of RVP-c is efficient.

\section{Discussion}

The proposed instruction fetch unit relies on the branch prediction mechanism being pipelined. If you want to achieve a high operating frequency when the branch prediction mechanism is not pipelined, you need to double the BTB entry width or prepare two BTBs. If it is not necessary to achieve a high operating frequency, it is possible to place an adder immediately after the BTB and use that value as a candidate for PC\_2. However, this path is likely to become a critical path.

If the RAM installed in the target FPGA has only one I/O port, it is necessary to divide the instruction memory into two banks like the gray instruction cache. {\bf The key idea of the proposed instruction fetch unit is that the value of the program counter and the value obtained by adding a constant to the value of the program counter can always be calculated in parallel as long as the RISC-V instruction set is used.} Therefore, the proposed instruction fetch unit can also be applied to the instruction cache architecture.

The performance improvement of RVP-c compared to VR-bp-c is reasonable, referring to Pollack’s law, which describes the relationship between hardware resources and processor performance. The empirical improvement, according to the law, is 20.2\%, which is proportional to the square root of the 44.6\% increase in slice usage compared to VR-bp-c. The obtained each performance improvement is bigger than this empirical improvement.

\section{Conclusion}

In this paper, we proposed the efficient instruction fetch unit supporting RISC-V compressed instructions and RVCoreP-32IC using this unit.

The most important point to consider when supporting the compressed instructions is that 32-bit instructions and 16-bit instructions coexist in the instruction memory and are arranged without gaps. By this point, an instruction fetch unit that supports the compressed instructions should be efficient to avoid a significant decrease in processor performance.

The proposed instruction fetch unit always fetches the entry of the instruction memory indicated by the program counter and the next entry in parallel. The proposed processor RVCoreP-32IC uses this unit to support RISC-V compressed instructions.

We compared the proposed processor RVCoreP-32IC with related research VexRiscv in Verilog HDL simulation and implementation on FPGA. Using the proposed instruction fetch unit, DMIPS, CoreMark value, and Embench value of RVCoreP-32IC achieved 42.5\%, 41.1\%, and 21.3\% higher performance than VexRiscv, respectively.

\bibliographystyle{IEEEtran}
\bibliography{main}

\vspace{12pt}

\end{document}